# A Novel Defence Scheme against Selfish Node attack in MANET

Gaurav Soni[1] and Kamlesh Chandrawanshi[2]

[1]Dept. of Computer Science & Engineering, TIT College, Bhopal, M.P, INDIA
`gauravsoni.rits@gmail.com`
[2]Dept. of Computer Science & Engineering, BIST College, Bhopal, M.P, INDIA
`kamlesh.vjti@gmail.com`

## ABSTRACT

*Security is one of the major issue in wired and wireless network but due to the presence of centralized administration not difficult to find out misbehavior in network other than in Mobile Ad hoc Network due to the absence of centralized management and frequently changes in topology security is one of a major issue in MANET. Only prevention methods for attack are not enough. In this paper a new Intrusion Detection System (IDS) algorithm has proposed against selfish node attack in MANET. Here the behavior of selfish node is unnecessary flooding the information in network and block all types of packets transferring between the reliable nodes. Proposed IDS Algorithm identifies the behavior of selfish node and also blocked their misbehavior activities. In case of selfish node attack network performance is almost negligible but after applying IDS on attack network performance is enhanced up to 92% and provides 0% Infection rate from attack.*

## 1. INTRODUCTION

A mobile ad hoc network (MANET) is a collection of mobile devices that can communicate with each other without the use of a predefined infrastructure or centralized administration. In addition to freedom of mobility, a MANET can be created quickly at a low cost, as it does not rely on existing network infrastructure. Due to this suppleness, a MANET is very useful for applications such as disaster relief, emergency operations, military service, maritime communications, vehicle networks, business meetings, site networks, robot networks, and so on. In these networks, besides acting as a host, each node also acts as a router and forwards packets to the correct node in the network once a route is established nodes are able to transfer their information to other nodes. To support this connectivity nodes use routing protocols such as Proactive routing protocols and Reactive routing protocols. Proactive routing protocols such as Destination Sequence Distance Vector (DSDV) protocol [1], nodes obtain routes by periodic exchange of topology information in the foam of maintaining the routing table. Reactive routing protocols such as AODV (Ad hoc On-Demand Distance Vector) [2] are ad hoc on demand routing protocols here nodes find out routes if requires. In the route discovery process of AODV protocol, intermediate nodes are responsible to connect a fresh path to the destination, sending discovery packets i.e RREQ (Route request) packets to the neighbor nodes and the number of nodes that are in radio range of sender will immediately respond to send back RREP (Route Reply) messages in network. If any the number of nodes are that not reach to destination because of some limitations like node moves out of range then in that case RERR (Route error) messages are generated. These error messages confirm the possibilities of best route selection with minimum hop counts. In addition, AODV enables intermediate nodes that have sufficiently fresh routes (with destination sequence number equal or greater than the one in the RREQ) to generate and send an RREP to the source node. Malicious node abuse this process and they instantly respond to the source node with fake information as though they have a fresh enough path to the destination. Therefore source node sends its data packets via this malicious node assuming it is a true path. Selfish node behavior may also be due to damaged nodes dropping packets unintentionally. In any case, the end result of the presence of a selfish node in the network is lost packets (both routing as well as data). In our study, we simulated selfish node attacks in wireless ad hoc networks and evaluated their effects on the network performance.

The paper organization is as follows: section 2 describes selfish node attack and related works are described in section 3. Proposed Algorithm is described in section 4. Network simulation results are presented in section 5 followed by conclusions in section 6.

## 2. SELFISH NODE ATTACK

Routing protocols are exposed to a variety of attacks. Selfish node attack is one such attack in which a malicious node doing a routing misbehavior in the route discovery packets of the routing protocol to advertise itself as having the shortest path to the node whose packets it wants to intercept [3,4]. This attacks aims at modifying the routing protocol so that traffic flows through a specific node controlled by the attackers. During the route discovery process, the source node sends route discovery packets to the intermediate nodes to find fresh path to the intended destination. Malicious nodes respond immediately to the source node as these nodes do not refer the routing table and drop all the routing packets and also flooding the false information of shortest route in network by that the number of nodes that are in radio range directly or indirectly forwarded the routing as well as data packets in the network. The source node assumes that the route discovery process is complete, ignores other route reply messages from other nodes and selects the path through the malicious node to route the data packets. The malicious nodes do this by assigning a high sequence number to the reply packet. In an ad-hoc network that uses the AODV protocol, a Selfish node absorbs the network traffic and drops all packets. To explain the Selfish Node we added a malicious node that exhibits Selfish behavior and capture the UDP packet and block the TCP packet or can't forward the TCP data to actual destination.

In a Selfish Node, after a while, the sending node understands that there is a link error because the receiving node does not send TCP ACK packets. If it sends out new TCP data packets and discovers a new route for the destination, the selfish node still manages to mislead the sending node. If the sending node sends out UDP data packets the problem is not detected because the UDP data connections do not wait for the ACK packets.

## 3. RELATED WORK

Standard Recently, a lot of research has focused on the cooperation issue in MANET. Several related issues are briefly presented here.

Khairul Azmi et al [5] present a new mechanism to detect selfish node. Each node is expected to contribute to the network on the continual basis within a time frame. Those which fail will undergo a test for their suspicious behavior. This scheme is also a based on monitor node. A monitoring node hears a request from its neighbouring node to forward a data packet; it will first check the time difference between *last request* and *last action* and status of the requestor.

Performance metrics are not measures in this paper now in present work we include the infection ratio and performance metrics. Future work of [5] like acknowledgement detail and their loss are also measure.

Al Shurman et al [6] have proposed two different solutions for black hole. The first solution suggests unicasting a ping packet from source to destination through multiple routes and then chooses a safe route based on the acknowledgement received. The second solution is based on keeping track of sequence numbers. But these solutions have a longer delay and lower number of verified routes.

Misbehavior detection and reaction are described in [7], by Marti, Giuli, Lai and Baker. The paper presents two extensions to the DSR algorithm: the watchdog and the path rater. The watchdog identifies misbehaving nodes by listening promiscuously to the next node transmission but not detect misbehavior in presence of ambiguous collisions, receiver collisions, limited transmission power, false misbehavior and partial dropping.

This technique is imperfect due to collisions, limited transmit power and partial dropping. Buchegger and Le Boudec [8] present the CONFIDANT protocol. Each node monitor the behaviour of its next hop neighbors in a similar manner to watchdog. Deciding the criteria for maintaining the friends list by Trust Manager is difficult.

CORE (Collaborative Reputation) [9] is a reputation based system proposed by Michiardi et al similar to CONFIDANT. The limitation with CORE is that the most reputed nodes may become congested as most of the routes are likely to pass through them. Also the limitations of the monitoring system in networks with limited transmission power and directional antennas have not been addressed in CORE.

Patcha et al [10] have proposed a collaborative architecture for black hole prevention as an extension to the watchdog method.

Bansal et al [11] have proposed a protocol called OCEAN (Observation-based Cooperation Enforcement in Ad hoc Networks), which is the enhanced version of DSR protocol. OCEAN uses a monitoring system and a reputation system to identify malicious nodes. But OCEAN fails to deal with misbehaving nodes properly. These papers have addressed the black hole attack problem on unicast routing protocols.

Balakrishnan [12] has proposed a TWOACK scheme which can be implemented as an add-on to any source routing protocol. Instead of detecting particular misbehaving node, TWOACK scheme detects misbehaving link and then seeks to alleviate the problem of routing misbehavior by notifying the routing protocol to avoid them in future routes. It is done by sending back a TWOACK packet on successful reception of every data packet, which is assigned a fixed route of two hops in the direction opposite to that of data packets. Basic drawback of this scheme includes it cannot distinguish exactly which particular node is misbehaving node. Sometime well behaving nodes became part of misbehaving link and therefore cannot be further used the network. Thus a lot of well behaved node may be avoided by network which results in losing of well behaved routes.

Vijaya [13] proposed another acknowledgement based scheme similar to TWOACK scheme This scheme detects the misbehaving link, eliminate it and choose the other path for transmitting the data. The main idea is to send 2ACK packet which is assigned a fixed route of two hops back in the opposite direction of the data traffic route and to reduce the additional routing overhead, a fraction of the data packets will be acknowledged via a 2ACK packet. This scheme also consists of multicasting method by which sender can broadcast information of misbehaving nodes so that other nodes can avoid path containing misbehaving nodes and take another path for the data transmission. Although routing overhead caused by transmission of acknowledgement packets is minimized but this scheme also suffers to detect the particular misbehaving node.

Usha and Radha [14] proposed Ack is better than TWOACK scheme, in which each send a normal Ack to its immediate source node after receipt of any kind of packet. This scheme requires an end to end Ack packet (i.e. Nack) to be sent between the source and the destination. Possible drawback includes lot of routing overhead because of Ack and Nack packets. Also due to nodes mobility probability of Nack packet reaching to source becomes smaller with the large number of intermediate nodes between source and destination.

Zeshan [15] proposed a two-fold approach for detection and isolation of nodes that drops data packets. First approach attempts to detect the misbehavior of nodes and will identify the malicious activity in network. It is done by sending an ACK packet by each intermediate node to its source node for confirming the successful reception of data packets. Other approach identifies exactly which intermediate node is doing malicious activity. It is done by monitoring the intermediate nodes of active route by the nodes near to active path which lies in their transmission range and by the nodes which are on the active route. When number of dropped packets by a particular node exceeds certain threshold, the monitoring node in that range declares that node as misbehaving node and broadcast this information in the network by that all the normal nodes are aware about the attacker. Main disadvantage of this scheme includes the overhead due to transmissions of acknowledgement packets by every intermediate node to the source and working of all nodes in promiscuous mode.

## 4. PROPOSED ALGORITHM

### 4.1. Algorithm for Selfish Node Creation and prevention

```
Set mobile node = M        //Total Mobile Nodes
Set source node = S        //S   M
Set Destination Node = D   // D   M
Set Routing Protocol =AODV
Start simulation time = t₀
Set radio range = rr;      //initialize radio range

RREQ_B(S, D, rr ) // broadcast for communication and send request packet to D node
        {
If ((rr<=250) && (next hop >0))
        {
                Compute route ()
                {
   rtable->insert(rtable->rt_nexthop); //nexthop to RREQ source
        if (dest==true)
{       send ack to source node with rtable;
Data_packet_send(s_no, nexthop, type)
                }
else            {
                destination not found;
                }
}
}
else { destination un-reachable ;
```

```
    }
}

Selfish_Node ()   //Selfish Node Work
        {
Check (incoming packet)
{ If (pkt == 'Routing')
{Capture and updated destination field ;
Send route ACK to sender;
}
Else if (pkt == 'TCP')
{Block TCP packet }
Else If {pkt =='UDP'}
{Capture UDP packet;
Can't Send to Destination;
}
Else ( pkt =='other')
{Drop; }
Set     (false_pkt= (scan_rate * pkts_max_) / selfish node);  // false packet send's to all normal
node

Selfish_Broadcast (inf_pkt, nexthop)
{
Set priority = 1                  // Higher priority
Send false_pkt = 100 pkts/ms            // greater than the limit
Find (number of pkt accepted node)
}
```

## 4.2. IDS for Elimination of Selfishness Algorithm

```
Set IDS node = p ; // IDS node
Set routing =AODV ;
RREQ_B(p, n, rr)        // broadcast for communication and send request packet to all n nodes
                         via p node
{
If ((rr<=250) && (next hop >0))
{
Set false_pks_rm (removal) = ( scan_rate *pkt s_max_ / selfish node);  // false packet disable
module
If (false_rm  => 100)
{ Selfish Node Block ; }
```

**Check_Selfishness (S,D,M)**

```
{
If ((node €M) && (pkt < 100 pkts/ms)
{
pkt accepted by neighbor;
pkt_Accept_limit();
}
Else { Node_Selfish()
{ can't accept by neighbor ;
Block pkts sender ;
```

```
}}}
If (any node belong is in radio range && receives that request packet && heavy load node)
{   Node l = week node   // 1   n
     Check load of  Node l ;
If (load > normal  load)
{   false_pkts_rm = 1;
{ Node infection remove via false_pkts_rm  parameter ;
}}
}
Node unreachable;
}
Node out of range;
}
```

### 4.3. Algorithm Description.

Behavior of selfish node is to be defined in first point of this section this attack drops all the packets in the network routing, TCP and UDP. The selfish node behavior are also explained in the foam of generating false packets in the network. The scan rate is calculated with maximum number of packets suppose number of nodes having capability of packets handling are 10 and scan rate is that capability of nodes in terms of time means in seconds or microseconds then suppose this is 1000 pktsps now divide the number of value of false packets by number of selfish node in the network. The value comes after calculation is the number of false packets in actual generate by each selfish node in the network.

An intrusion-detection system (IDS) can be defined as the tools, methods, and resources to help identify, assess, and report unauthorized or unapproved network activity. Intrusion detection is typically one part of an overall protection system that is installed around a system or device—it is not a stand-alone protection measure. In our simulation module we apply IDS module that protect through the selfish node behavior if Selfish node in the range of IDS. Very first IDS check which node update the routing table and send higher sequence number to the sender node, if find out so IDS sends the message to the sender node for elimination of that particular path where belongs selfish node and search new route according to IDS instruction. Here IDS internal module provides only protection of misbehave and provide trust communication between sender and destination. After prevention we detect selfish node via trace analysis and provide secure communication in MANET.

In our approach we have inbuilt IDS module with AODV routing and Selfish behavior module. Very first we attach IDS and Selfish module in the NS-2 package and update the make file through following command:

**selfish/selfish_logs.o selfish/selfish.o selfish/selfish_rtable.o**
**selfish/selfish_rqueue.o\idsaodv/idsaodv_logs.oidsaodv/idsaodv_rtable.o**
**idsaodv/idsaodv_rqueue.o idsaodv/idsaodv.o**

After that we update the packet.h file through PT_idsAODV= "IDSAODV"; and PT_selfish= "Selfish"; and compile the internal module if new object file generated then we create TCL (tool command language script) for the scenario creation and create the MANET scenario, TCL invoke the new module Selfish and IDS module and gives the behavior according to selfish and IDS module. Then we create two different type of Output file name as .tr (trace file) and .nam (network animator file) through TCL script. Trace file contain each information in particular discrete event of simulation and that file passes to awk (abstract window tool kit) and get the output in the form of routing overhead, throughput, average end-to-end delay etc.

Here we create three module names as AODV simple routing, IDS (intrusion detection and prevention system) module and selfish module step by step

## 5. SIMULATION ENVIRONMENT

The detailed simulation model is based on network simulator-2 (ver-2.31) [16], is used in the evaluation. The NS instructions can be used to define the topology structure of the network and the motion mode of the nodes, to configure the service source and the receiver to create the statistical data trace file and so on. Performance metrics are calculated from trace file (.tr), that has contained the all simulation information.

### 5.1. Simulation Parameters for Case Study.

In our scenario we take 30 nodes in which nodes 1-27 are simple nodes, and node 29 is a malicious nodes or Selfish node and node 28 is an IDS node. The simulation is done using ns-2, to analyze the performance of the network by varying the nodes mobility. The evaluated performances are given below. We are taking the following parameters for case study shown in table 1. Note down selfish nodes are consider one in case of attack to visualized the effect of attack but after applying IDS scheme consider the selfish nodes are two (29 and 14) to see the secure effect of IDS in network.

Table 1. Simulation Parameters for Case Study.

| Number of nodes | 30 |
|---|---|
| Selfish node | 1 or 2 |
| IDS Node | 1 |
| Dimension of simulated area | 800×600 |
| Routing Protocol | AODV |
| Simulation time (seconds) | 100 |
| Transmission Range | 250m |
| Traffic type | CBR |
| Packet size (bytes) | 512 |
| Number of traffic connections (TCP or UDP) | 20 |
| Maximum Speed (m/s) | 30 |

### 5.2. The Performance Metrics

In this paper we focus on evaluating the protocols under Selfish node or malicious nodes attack and measure the network performance after applying intrusion detection system with following criteria [2, 3, 8, 9, and 13].

1. *Packet Delivery Fraction (PDF):* The ratio of data the delivered to the destination to the data send out by source.
2. *Throughput:* Numbers of packets send or received in per unit of time.
3. *Normalized routing overhead:* This is the ratio of routing-related transmissions (RREQ, RREP, RERR etc) to data transmissions in a simulation. .

International Journal on Computational Sciences & Applications (IJCSA) Vol.3, No.3, June 2013

4. *End to End Delay*:   The difference in the time it takes for a sent packet and to reach the destination.

## 5.3. Results

In this section we present a set of simulation experiments to evaluate this protocol by comparing with the original AODV [5]

### 5.3.1. PDF Analysis in case of Attack and IDS

In this graph we observe that in case of attack only at time about 5 sec the PDF value is 65 but after that the PDF percentage continuously decreases and reach up to 15 %. But after applying IDS The PDF continuously increases w.r.t to time  it means that  IDS node will block the misbehavior activity of selfish node  and also providing the shortest and secure path to nodes by that they the PDF is about 92% in presence of IDS. This one is the important parameter to measure the network performance in network layer because if the number of packets is successfully delivered in the network then definitely the packet loss is minimum. By applying IDS we recover about 75% data delivery in network; this one is huge loss in attack case.

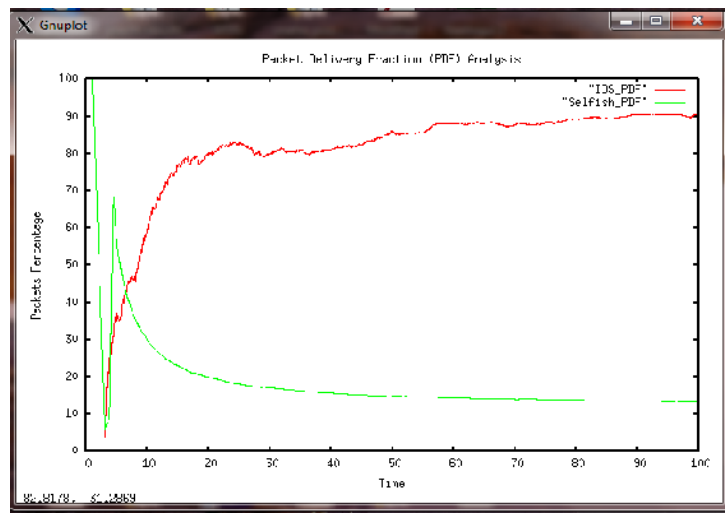

Figure1. PDF analysis in case of attack and IDS

### 5.3.2. Normalized Routing Overhead Analysis

This graph has show the routing load analysis in case of attack and IDS. Here we clearly observe the effect of attack in network. In case of attack only about 1000 packets are delivered because remaining packets are drop by selfish node then in case of attack NRL is minimum. But after applying IDS a normal secure routing packets delivered in network and also deliver successfully data between the nodes. Now in case of a IDS about 6000 routing packets are delivered, this is about 6 times higher than attack.

58

International Journal on Computational Sciences & Applications (IJCSA) Vol.3, No.3, June 2013

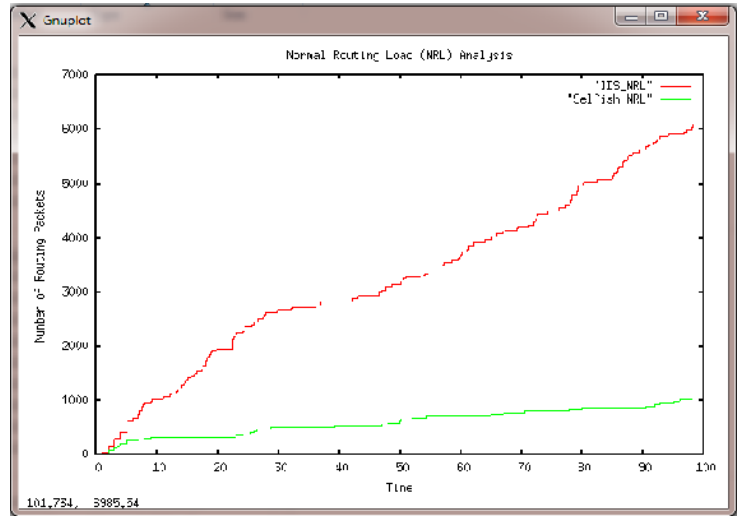

Figure 2. NRL analysis in case of attack and IDS

### 5.3.3. Throughput Analysis

Throughput analysis in in case of attack clearly show that the throughput in case of attack is negligible but after applying IDS number of packets send in network are continuously increases by that throughput are increases. Here we see that at the time about 22 seconds throughput is negligible in case of attack and IDS but after that in IDS case throughput increases up to 42 packets are send in per unit of time. Now in IDS case throughput is about 100% as compare to attack. The throughput value in graph is not visualized because it is less than 1.

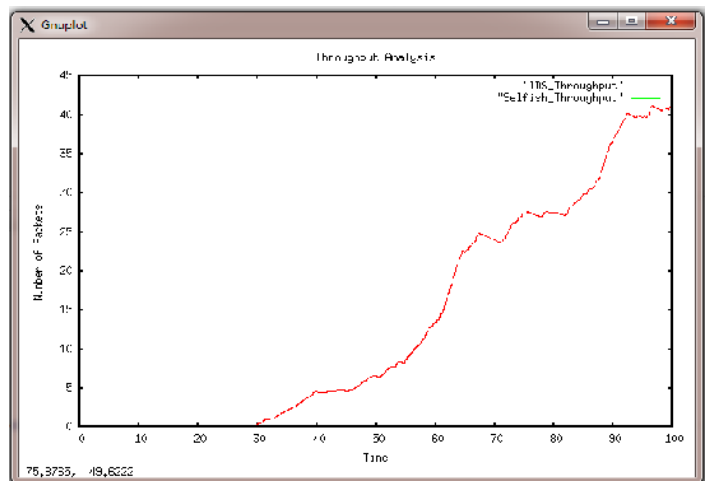

Figure 3. Throughput analysis in case of attack and IDS

### 5.3.4. Infection Percentage

Infection percentage represents the infection percentage w.r.t time. Infection percetage in case of attack are continuously increases reach up to 42%. At time about after 4 sec. the infection are in maximum percentage value but at the time of IDS the infection percentage is zero and not a

International Journal on Computational Sciences & Applications (IJCSA) Vol.3, No.3, June 2013

single packet is affected by selfish node attack.IDS will block the whole activity of selfish node and remove the infection from network.

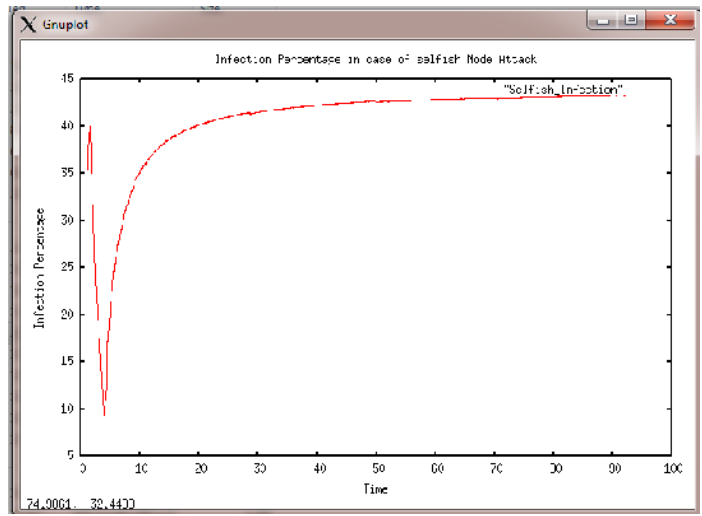

Figure 4. Infection Percentage in case of attack and IDS

Table 2 represents the effect of attack in TCP packets. Here the node 29 are drop the highest number of packets and remaining nodes are not deliver actual number of packets in network because most of the route request packets dropped by selfish node.

Table 2. TCP Packet Analysis in Case of Attack

| Sender Node | Packets Sends | Receiver Node | Packets Receives | Packets Drop by Node | Drop Packets |
|---|---|---|---|---|---|
| 0 | 19 | 7 | 8 | 29 | **67** |
| 2 | 27 | 11 | 6 | | |
| 5 | 7 | 19 | 3 | | |
| 6 | 6 | 21 | 1 | | |
| 22 | 20 | 24 | 1 | | |
| 25 | 7 | - | | | |
| | **Packets send = 86** | | **Packets receive = 19** | | |

TCP acknowledgement (Ack) details are given in table 3. Here it is clear that a negligible number of Ack are received by sender nodes. Now attacker drops about all the Ack packets and Ack drop by node field is nil it means that Ack are lost by attacker. There is no information in network which nodes are drop the Ack packets.

International Journal on Computational Sciences & Applications (IJCSA) Vol.3, No.3, June 2013

Table 3. TCP Acknowledgement Packet Analysis in Case of Attack

| Ack receiver Node | Ack packets receives | Ack drop by Node | Ack Drop |
|---|---|---|---|
| 0 | 3 | - | - |
| 2 | 6 | | |
| 5 | 1 | | |
| 22 | 8 | | |
| 25 | 1 | | |
| **Total Ack. receives** | **19** | | |

IDS scheme really show the actual performance of network which is shown in table 4.This table clearly show how many data packets are transfer, receive and drop in between sender, receiver and intermediate nodes.

Table 4. TCP Packet Analysis in Case of IDS

| Sender Node | Packets Sends | Receiver Node | Packets Receives | Packets Drop by Node | Drop Packets |
|---|---|---|---|---|---|
| 0 | 729 | 7 | 1415 | 0 | **9** |
| 2 | 414 | 11 | 401 | 1 | 10 |
| 5 | 161 | 19 | 306 | 2 | 9 |
| 6 | 310 | 21 | 695 | 5 | 3 |
| 22 | 1458 | 24 | 28 | 6 | 10 |
| 25 | 42 | 28 | 147 | 8 | 4 |
| | | | | 10 | 10 |
| | | | | 13 | 1 |
| | | | | 14 | 2 |
| | | | | 17 | 2 |
| | | | | 22 | 37 |
| | | | | 23 | 3 |
| | | | | 25 | 18 |
| | | | | 26 | 1 |
| | **Packets send = 3114** | | **Packets receive = 2992** | **Packet Drop** | **119** |

The table 5 is showing the actual information about the Ack. packets send by receiver and intermediate nodes in network. As compare to attack case here the almost all Ack. packets are receive. In both Ack. tables there is no information about attacker node because their behaviors are not normal as other node.

Table 5. TCP Acknowledgement Packet Analysis in Case of IDS

| Ack Node | Ack Packets Receives | Ack drop by Node | Ack Drop |
|---|---|---|---|
| 0 | 653 | 0 | 9 |
| 2 | 352 | 1 | 8 |
| 5 | 124 | 2 | 1 |
| 6 | 289 | 7 | 23 |
| 22 | 1385 | 8 | 2 |
| 25 | 19 | 11 | 36 |
|  |  | 12 | 11 |
|  |  | 17 | 6 |
|  |  | 19 | 33 |
|  |  | 20 | 9 |
|  |  | 23 | 5 |
|  |  | 24 | 12 |
|  |  | 25 | 2 |
|  |  | 26 | 1 |
|  |  | 28 | 9 |
| Total ACK Receives | 2822 | ACK. Drop | 167 |

## 6. CONCLUSION AND FUTURE WORK

Finally after visualize the results, it can be concluded that the Selfish node effect on performance metrics. Effect on packet loss is clearly visualized in PDF and throughput. As malicious node is the main security threat that effect the performance of the AODV routing protocol. Its detection is the main matter of concern. The acknowledgement(ACK) of TCP represents that due to fake information in network most of the senders are not obtain the ACK from receiver means all the ACK are lost and after applying IDS scheme every node that take part in routing will show the information of ACK packets.

Therefore the proposed IDS scheme work will be excellent to detect and defense the network from Selfish node attack. In Future we also detect the effect of selfish attack in performance matrices and also Selfish node for AODV can be implemented in real life scenario and its analysis can be compared with the analysis results.

International Journal on Computational Sciences & Applications (IJCSA) Vol.3, No.3, June 2013

**Authors**

**1)   Gaurav Soni** He is working as Assistant Professor in Technocrats Institute of Technology Engineering College, Bhopal, Madhya Pradesh, (M.P), India. He did his Bachelor degree in Computer Science. He finished his Master of Engineering in Computer Science department from RITS, Bhopal. Their area of interest is Mobile Ad hoc Network. He is Attended four International Conferences and one national conference & Published one paper in international conference.

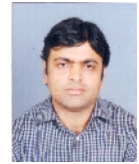

**2) Kamlesh Chandrawanshi** He is working as Assistant Professor in Bansal Institute of Science & Technology Engineering College, Bhopal, Madhya Pradesh (M.P), India. He did his Bachelor degree in Computer Science. He finished his Master of Engineering in Computer Science from VJTI, Mumbai. Their area of interest is Wireless Network & Sensor Network. He is published seven research papers in international conference & Published one paper in national conference.

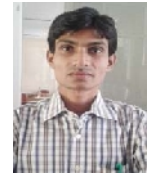